\documentclass[RNAAS]{aastex63}

\graphicspath{{./}{figures/}}
\begin{document}

\title{Simultaneous observations of the northern TESS sectors by the Zwicky Transient Facility}

\correspondingauthor{Jan van Roestel}
\email{jvanroes@caltech.edu}

\author{Jan van Roestel}
\affiliation{Division of Physics, Math, and Astronomy, California Institute of Technology, Pasadena, CA 91125, USA}

\author{Eric C. Bellm}
\affiliation{DIRAC Institute, Department of Astronomy, University of Washington, 3910 15th Avenue NE, Seattle, WA 98195, USA}

\author{Dmitry A. Duev}
\affiliation{Division of Physics, Math, and Astronomy, California Institute of Technology, Pasadena, CA 91125, USA}

\author{Christoffer Fremling}
\affiliation{Division of Physics, Math, and Astronomy, California Institute of Technology, Pasadena, CA 91125, USA}

\author{Matthew J. Graham}
\affiliation{Division of Physics, Math, and Astronomy, California Institute of Technology, Pasadena, CA 91125, USA}

\author{Frank Masci}
\affiliation{Infrared Processing and Analysis Center, California Institute of Technology, Pasadena, CA 91125, USA}

\author{Lin Yan}
\affiliation{Division of Physics, Math, and Astronomy, California Institute of Technology, Pasadena, CA 91125, USA}

\author{Daniel A. Goldstein}
\affiliation{Division of Physics, Math, and Astronomy, California Institute of Technology, Pasadena, CA 91125, USA}

\author{Michael Medford}
\affiliation{Astronomy Department, University of California, Berkeley, CA 94720, USA}
\affiliation{Lawrence Berkeley National Laboratory, Berkeley, CA 94720, USA}
\affiliation{Lawrence Livermore National Laboratory, Livermore, CA 94550, USA}

\author{Charlotte A. Ward}
\affiliation{Lawrence Berkeley National Laboratory, Berkeley, CA 94720, USA}
\affiliation{Department of Astronomy, University of Maryland, College Park, MD 20742, USA.}

\author{S. R. Kulkarni}
\affiliation{Division of Physics, Math, and Astronomy, California Institute of Technology, Pasadena, CA 91125, USA}

\author{Thomas A. Prince}
\affiliation{Division of Physics, Math, and Astronomy, California Institute of Technology, Pasadena, CA 91125, USA}


\section{} 
The Transiting Exoplanet Survey Satellite (TESS) \citep{tess} is a powerful facility for studying a broad range of astrophysical objects. The Zwicky Transient Facility (ZTF) \citep{bellm+19,2019PASP..131g8001G,masci+19} is conducting a nightly public survey of all 13 TESS northern sectors in 2019-2020. ZTF will observe the portions of the current TESS sector visible from Palomar Observatory each night. Each ZTF pointing will have one exposure each with $g$ and $r$ filters, totaling two images per night. The first northern sector, Sector 14, was observed from July 18 -- August 15, 2019. The observations of the second northern sector, Sector 15, began on August 15, 2019. The majority of Sectors 14 and 15 have been covered by ZTF, except for a portion of TESS Camera 4, due to the visibility limits. ZTF is also making additional nightly $g$- and $r$-band observations of denser stellar regions (e.g. near the Galactic Plane) to better facilitate variability studies of Galactic objects.

\begin{figure}[h!]
\begin{center}
\includegraphics[width=0.95\textwidth,angle=0]{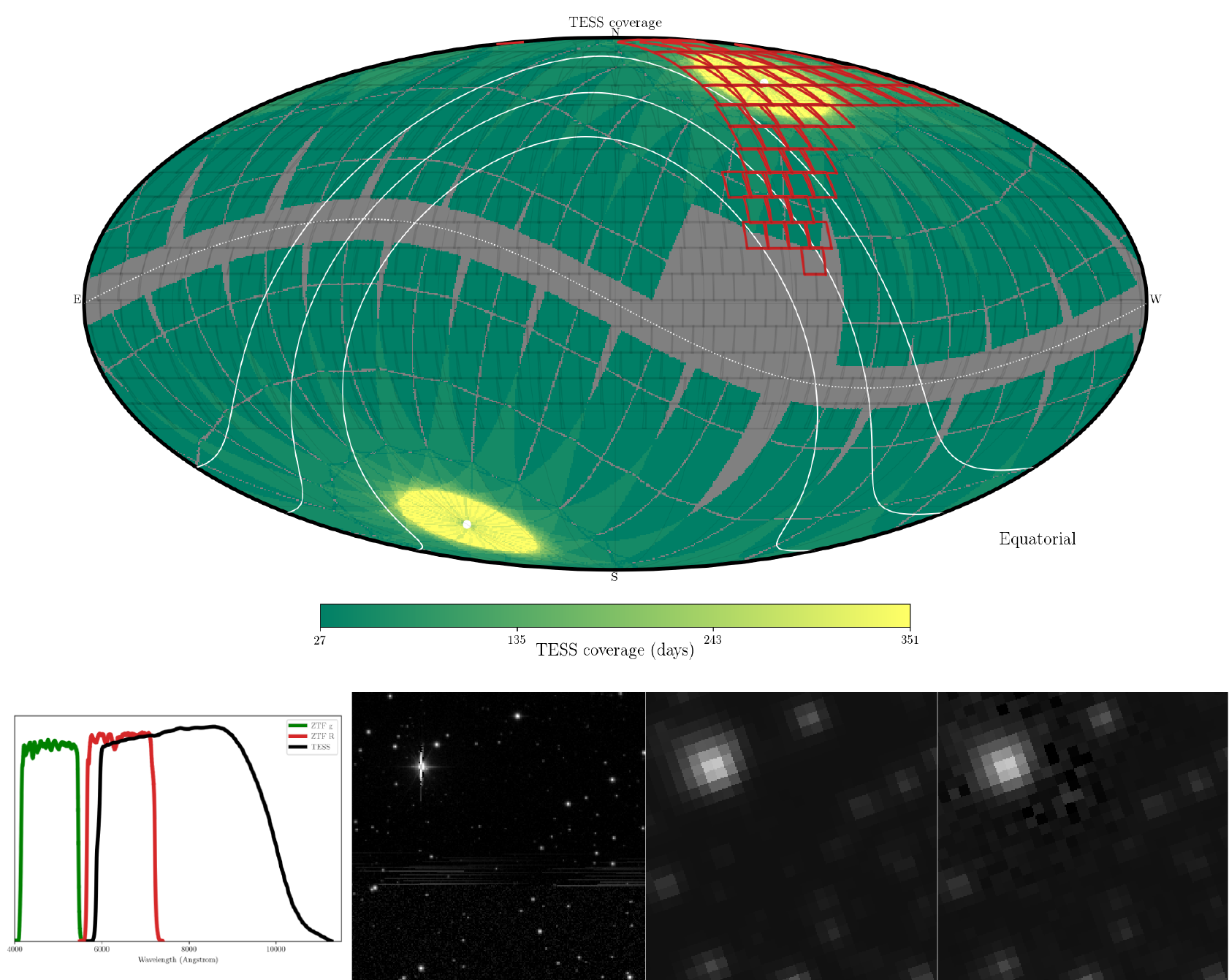}
\caption{\label{fig:1} The sky coverage of TESS for the first 2 years of observations. The color indicates how many days an area is observed by TESS. The ZTF-fields that cover TESS-sector 14 are over-plotted in red. In the lower left, the filter response curves are shown, green and red indicate ZTF $g$ and $r$, black shows the TESS bandpass. The pictures on the lower right show a ZTF image, a TESS image of the same area, and a simulated TESS images based on the ZTF image using Tractor.}
\end{center}
\end{figure}

ZTF has a FOV of 55.0 square degrees and a light-sensitive area of 47.7 square degrees within that FOV. A TESS sector can be completely covered with about 60 ZTF fields. ZTF has 1\arcsec\ pixels (2\arcsec\ PSF) compared to the TESS 21\arcsec\ pixels (50\% enclosed flux). Consequently, ZTF observations will allow identification of transient and variable sources within a TESS pixel, particularly important if more than one source contributes to the variable flux. In addition, ZTF goes deeper than TESS, having a median 5-$\sigma$ limiting magnitude of 20.6 in $r$-band and 20.8 in $g$-band. This allows the identification of weak sources that may be contributing to the TESS measured flux. See Figure \ref{fig:1} for an example of ZTF and TESS images of the same field. ZTF $g$- and $r$-band measurements will complement TESS 600-1000\,nm observations, allowing estimates of temperature and other spectral characteristics. See the comparison of spectral bands in Figure \ref{fig:1}. 
ZTF saturates between m$\sim$12-13. 

ZTF will release data from TESS fields in three forms:
in addition to the nightly alerts distributed by established ZTF brokers, nightly alerts converted to JSON format are distributed via ZTF's bucket on Google Cloud as a tarball \citep{jupyter}, and monthly photometric light curves also distributed via Google Cloud.
Because ZTF alerts are issued nightly, they can identify transient and variable sources that are active, allowing ground-based follow-up by the community on a prompt timescale, faster than the timescale on which TESS data is typically available. 
Initial results are provided in earlier ATELs \citep{2019ATel12952....1Y,2019ATel12959....1B}.

Details on how to access alerts and light curves are given in a Jupyter notebook accessible, e.g. via Google Colaboratory \citep{jupyter}. ZTF-TESS alerts are cross-matched against several external catalogs, including 2MASS PSC, AllWISE, IPHAS DR2 and Gaia DR2. Finally, we note that no filtering is applied to the ZTF-TESS alerts prior to their public release. We encourage spectroscopic followup of alert candidates as quickly as possible to maximize the potential science return from the ZTF-TESS observations.
In addition, supernova candidates identified using a combination of automated filters and human vetting by the ZTF collaboration are automatically posted to the Transient Name Server (TNS) \citep{tns} with the ``sender'' keyword set to ``ZTF\_TESS''. 

ZTF light curves for all objects observed concurrently by TESS and ZTF will be available publicly less than one month from the end of a TESS sector observation campaign. Details on how to access and filter the ZTF light curve data will be given in the same Jupyter notebook as mentioned above. More functionality to work with ZTF light curve data will also be available with subsequent light curve releases.

Tractor \citep{tractor} models of ZTF images can be used to simulate the overlapping TESS observations.  This technique exploits the higher resolution of ZTF to resolve the position and intensities of sources that may be blended in TESS images. The modelling software used to produce co-added ZTF images and simulated TESS images corresponding to the TESS fields is available on Github \citep{legacypipe_github, tractor_github}. Example code and detailed documentation for ZTF/TESS modelling will be disseminated in the coming months.

\acknowledgments

ZTF is a project led by PI S. R. Kulkarni at Caltech, and includes IPAC; WIS, Israel; OKC, Sweden; JSI/UMd, USA; UW, USA; DESY, Germany; NRC, Taiwan; UW Milwaukee, USA, and LANL USA. ZTF acknowledges the generous support of the NSF under AST MSIP Grant No 1440341. Alert distribution service provided by DIRAC@UW. Alert filtering is being undertaken by the GROWTH marshal system, supported by NSF PIRE grant 1545949. We acknowledge the support from the Heising-Simons Foundation under Grant No. 12540303. We acknowledge support from the University of California Office of the President for the UC Laboratory Fees Research Program In-Residence Graduate Fellowship (Grant ID: LGF-19-600357).


\begin{thebibliography}{}
\expandafter\ifx\csname natexlab\endcsname\relax\def\natexlab#1{#1}\fi
\providecommand{\url}[1]{\href{#1}{#1}}
\providecommand{\dodoi}[1]{doi:~\href{http://doi.org/#1}{\nolinkurl{#1}}}
\providecommand{\doeprint}[1]{\href{http://ascl.net/#1}{\nolinkurl{http://ascl.net/#1}}}
\providecommand{\doarXiv}[1]{\href{https://arxiv.org/abs/#1}{\nolinkurl{https://arxiv.org/abs/#1}}}

\end{thebibliography}


\begin{thebibliography}{99}
\bibitem[J Schliegel (2017)]{tess} J Schliegel, 2017, “TESS Observatory Guide, version 1.1”. In: TESS Guest Investigator Program: TESS Observatory Guide, Version 1.1 (June 30, 2017). TESS Science Support Center, NASA Goddard Space Flight Center, Greenbelt, MD. 
\bibitem[Bellm et al. (2019)]{bellm+19} Bellm et al., 2019, \pasp, 131, 018002.
\bibitem[Graham et al. (2019)]{2019PASP..131g8001G} Graham et al., 2019, \pasp, 131, 078001.
\bibitem[Masci et al. (2019)]{masci+19}Masci, F.J. et al., 2019, \pasp, 131(995), p.018003.
\bibitem[Duev (2019)]{jupyter}Duev, 2019, \url{https://colab.research.google.com/github/dmitryduev/kowalski/blob/master/nb/tess.ipynb}.
\bibitem[Yan et al. (2019)]{2019ATel12952....1Y} Yan et al. 2019, ATel \#12952.
\bibitem[Burdge et al. (2019)]{2019ATel12959....1B} Burdge et al. 2019, ATel \#12959.
\bibitem[Transient Name Server (2016)]{tns} Transient Name Server, \url{https://wis-tns.weizmann.ac.il/}
\bibitem[Lang et al. (2016)]{tractor} Lang et al., 2016, Astrophysics Source Code Library, `The Tractor:Probabilistic astronomical source detection and measurement'. 
\bibitem[Lang (2019a)]{legacypipe_github}Lang, 2019, \url{https://github.com/legacysurvey/legacypipe}.
\bibitem[Lang (2019b)]{tractor_github}Lang, 2019, \url{https://github.com/dstndstn/tractor}.
\end{thebibliography}
\end{document}